\documentclass[prb,aps,reprint]{revtex4-1}

\usepackage{graphicx}

\usepackage{bm}
\usepackage[english]{babel}

\usepackage{amsmath}
\usepackage{SIunits}

\usepackage{epstopdf}

\begin{document}

\title{Suppression of contact-induced spin dephasing in graphene/MgO/Co spin-valve devices by successive oxygen treatments}

\author{F. Volmer}
\author{M. Dr\"{o}geler}
\author{E. Maynicke}
\author{N. von den Driesch}
\author{M. L. Boschen}
\author{G. G\"{u}ntherodt}
\author{C. Stampfer}
\author{B. Beschoten}
\thanks{E-mail address: bernd.beschoten@physik.rwth-aachen.de}
\affiliation{2nd Institute of Physics and JARA-FIT, RWTH Aachen University, D-52074 Aachen, Germany}

\date{\today}

\begin{abstract}
By successive oxygen treatments of graphene non-local spin-valve devices we achieve a gradual increase of the contact resistance area products ($R_cA$) of Co/MgO spin injection and detection electrodes and a transition from linear to non-linear characteristics in the respective differential d$V$-d$I$-curves. With this manipulation of the contacts both spin lifetime and amplitude of the spin signal can significantly be increased by a factor of seven in the same device. This demonstrates that contact-induced spin dephasing is the bottleneck for spin transport in graphene devices with small $R_cA$ values. With increasing $R_cA$ values, we furthermore observe the appearance of a second charge neutrality point (CNP) in gate dependent resistance measurements. Simultaneously, we observe a decrease of the gate voltage separation between the two CNPs. The strong enhancement of the spin transport properties as well as the changes in charge transport are explained by a gradual suppression of a Co/graphene interaction by improving the oxide barrier during oxygen treatment.
\end{abstract}

\maketitle

\section{introduction}

Most graphene-based spin transport devices exhibit spin diffusion lengths of several micrometers at room temperature. However, there is a huge sample-to-sample variation in the reported spin lifetimes which vary between tens of ps and several ns at room temperature.\cite{Tombros2007,PhysRevLett.107.047206,PhysRevLett.107.047207,PhysRevB.84.075453,doi:10.1021/nl301567n,doi:10.1021/nl301050a,PhysRevB.86.161416,Dlubak2012,abel:03D115,PhysRevB.87.081405,PhysRevB.87.081402,1882-0786-6-7-073001,Neumann2013,2014arXiv1406.2439D,2014arXiv1406.4656G} The predicted long intrinsic spin lifetimes of graphene are most likely masked by extrinsic sources of spin scattering and spin dephasing.\cite{PhysRevLett.103.146801,PhysRevB.80.041405,1367-2630-14-3-033015} In particular, the direct contact of graphene to the underlying wafer and the electronic properties of the deposited spin injection and detection barriers are discussed as key factors that limit spin transport.\cite{PhysRevLett.105.167202,Dlubak2012,doi:10.1021/nl2042497,PhysRevB.80.041405,PhysRevB.88.161405}

\begin{figure}[t]
	\includegraphics{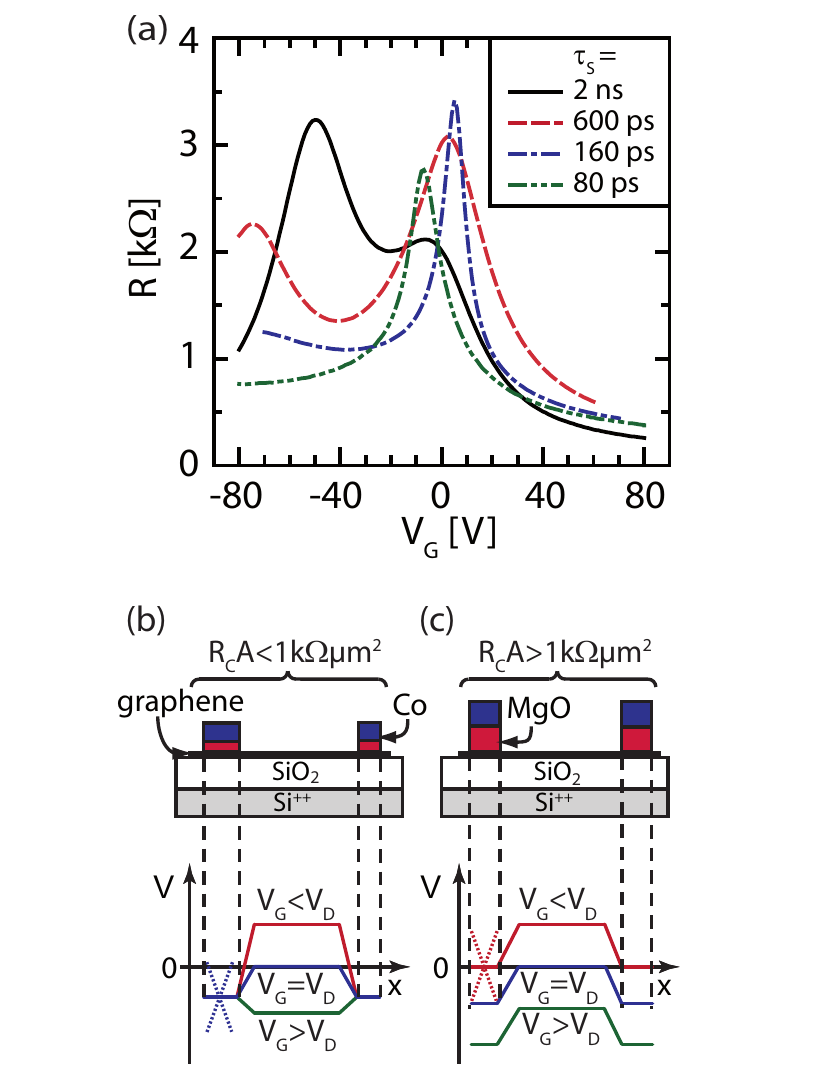}
	\caption{(Color online) (a) Gate dependent graphene resistance for selected spin transport devices with room temperature spin lifetimes ranging from 80~ps to 2~ns (data taken from Ref. \cite{PhysRevB.88.161405}). Devices with long spin lifetimes exhibit a 2nd charge neutrality point at negative gate voltages. The gate voltage separation of both charge neutrality points is smallest for the device with the longest spin lifetime of 2~ns. In addition to carrier doping of the underlying graphene layer, the interaction of the Co/MgO electrodes with graphene suppresses the electric field effect by the gate voltage $V_G$ which can result in (b) pinning (devices with $R_cA<\unit{1}{\kilo\ohm\micro\meter\squared}$) or (c) no pinning (devices with $R_cA>\unit{1}{\kilo\ohm\micro\meter\squared}$) of the chemical potential in graphene underneath the electrodes (corresponding Dirac cones are indicated by dashed lines).}
	\label{fig:fig1}
\end{figure}

Recently, we addressed the role of the contact resistance area product ($R_cA$) on spin transport in both single-layer graphene (SLG) and bilayer graphene (BLG) spin-valve devices where electron spins are injected into graphene from ferromagnetic Co electrodes across MgO barriers and are also detected by Co/MgO electrodes in non-local geometry.\cite{PhysRevB.88.161405} At room temperature, we only observed long spin lifetimes in devices with $R_cA>\unit{1}{\kilo\ohm\micro\meter\squared}$, which is consistent to other studies.\cite{PhysRevLett.105.167202,PhysRevB.86.235408} In our case, these devices also showed a second contact-induced charge neutrality point (CNP) in gate dependent charge transport at large negative gate voltages. By comparing many SLG and BLG spin transport devices we found the trend that the gate voltage separation between the two CNPs gets smaller in devices which exhibit longer spin lifetimes (see Fig.~\ref{fig:fig1}(a)).

We attributed the existence of the 2nd CNP in devices with $R_cA>\unit{1}{\kilo\ohm\micro\meter\squared}$ to the influence of the Co/MgO electrodes on the electronic states of the underlying graphene layer. Devices with thin MgO layers exhibit metallic pinholes ($R_cA<\unit{1}{\kilo\ohm\micro\meter\squared}$). These favor interactions at the contact between the metallic Co electrode and the graphene layer which results in Fermi level pinning in graphene underneath the electrodes.\cite{PhysRevB.79.245430} In these devices the back gate voltage can only control the graphene carrier density between the contacts while the graphene carrier density underneath the contacts is unaffected (Fig.~\ref{fig:fig1}(b)). In charge transport measurements these devices only show a single CNP and typical spin lifetimes on the order of 100~ps (see green curve in Fig.~\ref{fig:fig1}(a)). In contrast, we argue that a sufficiently thick and homogeneous MgO barrier prevents this charge carrier density pinning (see also Ref. \cite{nouchi:253503}). Therefore, the back gate voltage changes the carrier density in all graphene parts in these devices. The existence of the 2nd CNP at  negative gate voltages shows additional \textit{n}-doping by the electrodes ([Fig.~\ref{fig:fig1}(c) and further explanations in Sec.~\ref{oxidation_mechanism}).\cite{PhysRevLett.101.026803}

So far, both the increase of spin lifetime and the appearance of the second charge neutrality point with larger $R_cA$ values became only apparent by comparison of many devices both with small and high resistive contacts. However, such an observation is not unambiguous as spin and charge transport properties are most likely also sensitive to other extrinsic device properties that can show device-to-device variations. In this paper, we therefore explore the influence of Co/MgO spin injection/detection electrodes by repeated manipulation of the contact characteristics in the same device. The manipulation is conducted by multiple oxygen treatments after the initial device fabrication which results in an incremental increase of the effective oxide thickness at the graphene/MgO/Co interface. With this procedure we observe a complete transition from low resistive, transparent contacts ($R_cA<\unit{1}{\kilo\ohm\micro\meter\squared}$) to high resistive contacts that show non-linear differential \textit{I-V}-curves ($R_cA>\unit{1}{\kilo\ohm\micro\meter\squared}$). With this change of the contact characteristics, the spin transport properties can be enhanced significantly in the same device, i.e. the spin lifetime and the spin signal can be increased by a factor of 7. At the same time, we observe the appearance of the contact-induced 2nd CNP in charge transport which highlights the strong influence of the contacts on both spin and charge transport properties.

This paper is organized as follows: In the next section we describe the fabrication process and measurement techniques of the samples. The manipulation of the contact properties by exposing the devices to oxygen environments and its impact on the charge and spin transport are discussed in Sec.~\ref{oxygen_treatments}. In Sec.~\ref{oxygen_doping} we demonstrate that the increase in spin lifetime after oxygen treatments is not due to the oxygen doping of the graphene flake. The possible oxidization mechanism of the graphene/MgO/Co interface and the appearance of the second charge neutrality point are discussed in Sec.~\ref{oxidation_mechanism}. There we also propose a model explaining the strong increase of the spin lifetime which we observe near the contact-induced charge neutrality point. The conclusion is given in Sec.~\ref{conclusions}.

\section{sample fabrication and measurement techniques}
\label{fabrication}

\begin{figure}[tb]
	\includegraphics{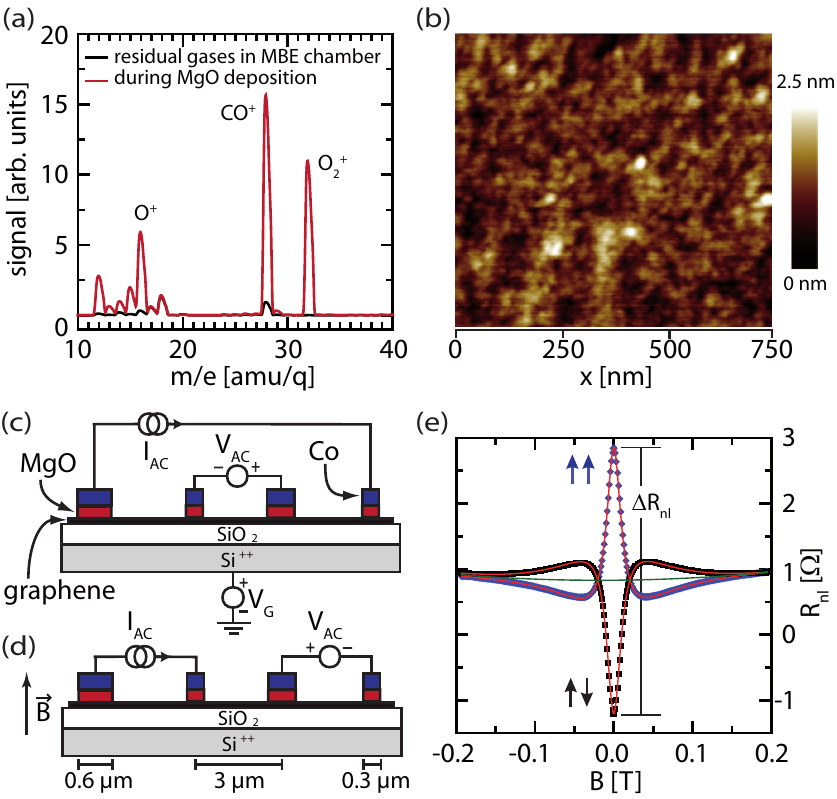}
	\caption{(Color online) (a) Residual gas analysis of the MBE chamber at base pressure (black curve) and during MgO deposition (red curve). The latter demonstrates the loss of (molecular) oxygen during MgO deposition which results in oxygen deficient MgO layers. (b) AFM image of a BLG sample after $\textrm{MgO}_x$ deposition: rms roughness of 0.4~nm and peak-to-peak roughness of 2~nm. (c) Schematic cross-section with wiring configuration for 4-terminal gate dependent ($V_G$) resistance measurements and (d) wiring for non-local spin transport measurements. (e) Hanle spin precession measurement of device B for a perpendicular magnetic field sweep with parallel and antiparallel alignments of the respective spin injection and detection electrodes.}
	\label{fig:fig2}
\end{figure}

We fabricated exfoliated SLG and BLG devices on $\textrm{SiO}_2$(\unit{300}{\nano\meter})/$\textrm{Si}^{++}$ wafers. After the deposition of the flake the wafers are put into acetone and thereafter into isopropyl alcohol to remove possible glue residuals. In the next step the e-beam lithography is carried out with PMMA dissolved in ethyl lactate and n-butyl acetate. The developer is a mixture of isopropyl alcohol and methyl isobutyl ketone. Prior to electrode deposition the samples are stored under UHV conditions in a molecular beam epitaxy (MBE) system for several days to allow for sufficient outgassing of the before-mentioned chemicals and water residuals. We use  electron-beam evaporation from MgO crystals (99.95\% metals basis) and Co pellets (99.95\% metals basis) at a base pressure of $\unit{1 \times \power{10}{-10}}{\milli\bbar}$. The deposition rates are $\unit{0.005}{\nano\meter\per\second}$ and $\unit{0.015}{\nano\meter\per\second}$ for MgO and Co, respectively, both at an acceleration voltage of \unit{4.5}{\kilo\volt}. We first grow the MgO spin injection/detection barrier with varying thicknesses up to 3~nm followed by 35-nm-thick ferromagnetic Co electrodes. During deposition the thicknesses of the evaporated layers are monitored by quartz oscillators. A residual gas analysis during MgO deposition shows the existence of both atomic and molecular oxygen [see red curve in Fig.~\ref{fig:fig2}(a)]. For comparison, we also included the gas analysis of the MBE chamber when no material is deposited as a reference [see black curve in Fig.~\ref{fig:fig2}(a)]. As the detected oxygen stems from the MgO crystals and is partially pumped out of the chamber, we conclude that the deposited MgO$_x$ layer is oxygen deficient ($x<1$).\cite{PhysRevB.73.205412,PhysRevLett.100.246803} On the other hand, oxygen vacancies in MgO tunneling barriers are known to disturb the tunneling process by introducing non-coherent tunneling channels.\cite{PhysRevLett.100.246803} We thus expect that oxygen vacancies in our oxygen deficient $\textrm{MgO}_x$ barriers may partially be healed by post-oxygen treatment.

In Fig.~\ref{fig:fig2}(b) we show an atomic force microscope (AFM) image of a 3-nm thick MgO layer which was deposited onto a SLG flake. Typical peak-to-peak values vary between 1.5 and 2~nm with rms values ranging between 0.3 and 0.5~nm \cite{PhysRevLett.107.047206}. All our MgO layers show island formation (Volmer-Weber growth). It is difficult to achieve layer-by-layer growth of MgO on graphene on an amorphous $\textrm{SiO}_2$ substrate as the graphene flake follows the corrugation of the substrate to some extent \cite{Ishigami2007,Cullen2010} and the graphene favors an overall high surface diffusion.\cite{wang:183107} Although it is reported that a wetting layer of $\textrm{TiO}_2$ can help to get atomically smooth MgO on graphene \cite{wang:183107} which can improve the tunneling characteristics of the contacts, it is, however, not obvious how the graphene-to-wetting-layer interface influences the spin properties after spin injection. We therefore restrict ourselves to not use any wetting layers in the present study. A further constraint for getting homogeneous tunneling barriers are almost unavoidable resist residues on top of the graphene flake after lithography and development\cite{doi:10.1021/nl203733r}, which in combination with the island growth mode may favor the formation of pinholes especially for thinner MgO layers.

A schematic cross-section of a spin-valve device is shown in Figs.~\ref{fig:fig2}(c) and (d). The center-to-center separation between neighboring electrodes is $\unit{3}{\micro\meter}$. If the size of the graphene flakes allows, we place the outermost electrodes at larger distances to minimize their influence on non-local spin measurements. The electrode widths alternate between $\unit{300}{\nano\meter}$ and $\unit{600}{\nano\meter}$ to achieve different coercive fields for magnetization reversal between neighboring Co electrodes. The highly \textit{p}-doped Si-wafer is used as a back gate which allows changing the charge carrier density $n=\alpha \left(V_{\textrm{G}} -V_{\textrm{CNP}}\right)$ in graphene according to the capacitor model~\cite{Novoselov2004} with $\alpha \approx \unit{7.18 \times \power{10}{10}}{\reciprocal\volt\centi\meter\rpsquared}$, $V_{\textrm{G}}$ being the applied gate voltage and $V_{\textrm{CNP}}$ being the gate voltage of the maximum resistance at the charge neutrality point. The gate dependent graphene resistance is measured in a local 4-terminal geometry (Fig.~\ref{fig:fig2}(c)).

Hanle spin precession measurements are performed in non-local 4-terminal geometry (Fig.~\ref{fig:fig2}(d)) with the external magnetic field $B$ applied in perpendicular direction.
All measurements are performed at room temperature (RT) under vacuum conditions at a base pressure of $\unit{6 \times \power{10}{-4}}{\milli\bbar}$. We use standard ac lock-in techniques, where the reference signal modulates the current through the device at a frequency of $\unit{18}{\hertz}$. The rms-values of the current are $\unit{1}{\micro\ampere}$ for gate dependent resistance measurements and $\unit{20}{\micro\ampere}$ for spin measurements.

Typical Hanle curves are depicted in Fig.~\ref{fig:fig2}(e) for device B (see next section) with both parallel and antiparallel alignments of the inner Co electrodes in Fig.~\ref{fig:fig2}(d). The non-local spin resistance $\Delta R_{\textrm{nl}}$ can easily be determined at $B=0$~T. The Hanle depolarization curves can be described by the steady-state Bloch-Torrey equation:\cite{ISI000249789600001,PhysRevB.37.5312}
\begin{equation}
	\frac{\partial \vec{s}}{\partial t}\;=\;\vec{s}\times \vec{\omega}_0+D_{\text{s}}\nabla^2\vec{s}-\frac{\vec{s}}{\tau_{\text{s}}}\;=0,
\end{equation}
where $\vec{s}$ is the net spin vector, $\vec{\omega_0}=g\mu_{B} \vec{B}/\hbar$ is the Larmor frequency where we set $g=2$, $D_{\text{s}}$ is the spin diffusion constant, and $\tau_{\text{s}}$ is the transverse spin lifetime. With $L$ being the distance between spin injection and spin detection electrodes, we define the following dimensionless parameters: $b\equiv g \mu_{\text{B}} B / \hbar \tau_{\text{s}}$, $l\equiv L/\sqrt{2 D_{\text{s}} \tau_{\text{s}}}$ and $f(b)=\sqrt{1+b^2}$. With a simplified analytical solution we use the following fit function to describe the Hanle curves:~\cite{ISI000249789600001,PhysRevB.37.5312}

\begin{align}
R_{\textrm{nl}}^{\textrm{Hanle}} = \Delta R_{\textrm{nl}}\; \frac{1}{2f(b)} \left[ \sqrt{1+f(b)}\; \text{cos}\left( \frac{l b}{\sqrt{1+f(b)}} \right) \right. \nonumber \\
\left. - \frac{b}{\sqrt{1+f(b)}}\; \text{sin}\left( \frac{l b}{\sqrt{1+f(b)}} \right)  \right] \text{exp}\left( -l \sqrt{1+f(b)} \right).
\end{align}

As expected, both Hanle depolarization curves merge at larger fields. However, they do not become constant but rather slightly increase above $|B|>0.2$~T. The increase results from a $B$ dependent background signal which we usually observe in all measurements. We account for this background by adding a polynomial of second order to our fit function:

\begin{equation*}
    R_{\textrm{nl}}^{\textrm{total}}(B) \;=\; R_{\textrm{nl}}^{\textrm{Hanle}}(B) + c_2 B^2 + c_1 B + c_0 \;.
\end{equation*}

Hanle fits are seen as red curves in Fig.~\ref{fig:fig2}(e). For illustration, we also include the parabolic background signal as a green curve.

\section{spin and charge transport after oxygen treatments}
\label{oxygen_treatments}

In the following section, we explore how subsequent oxygen treatments will influence the Co/MgO contact characteristics in our as-fabricated spin transport devices, i.e. the magnitude of the $R_cA$  products and the shape of the d$V$/d$I$ curves and, secondly, how they will change both spin and charge transport properties. We first focus on two representative devices both with 2-nm thick MgO barriers which result in low resistive transparent contacts in the as-fabricated devices due to the before-mentioned pinholes. Both devices (SLG device A and BLG device B) have identical electrode widths and separations (see also Fig.~\ref{fig:fig5}(a) for BLG device B). We note that the observed differences by oxygen treatments are not related to the number of graphene layers. This notion is supported by two additional devices (C and D, both SLG) which will be discussed in the following sections.

\subsection{SLG device}

\begin{figure}[tb]
	\includegraphics{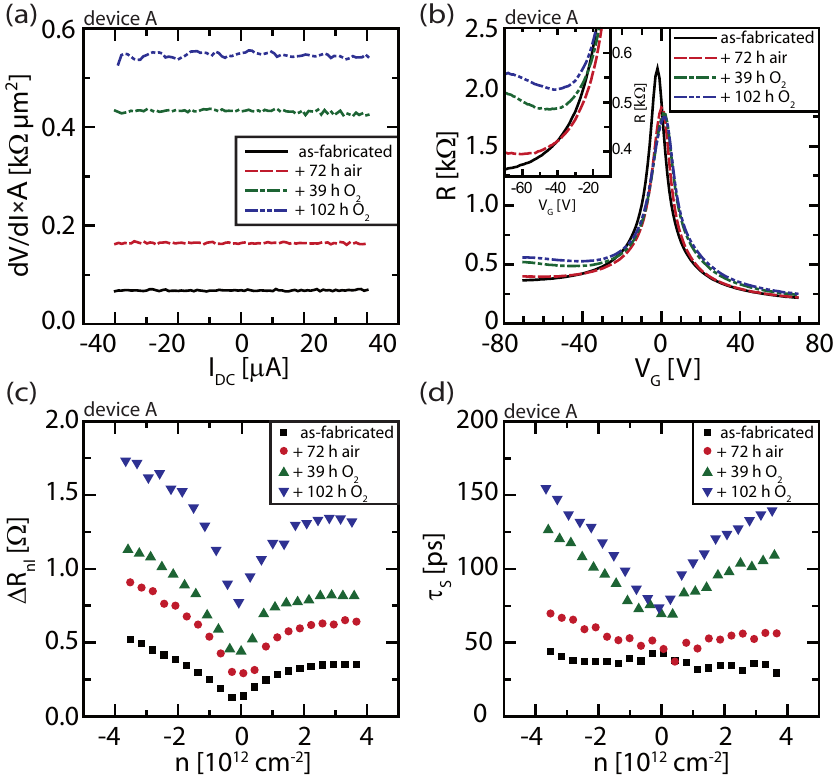}
	\caption{(Color online) Room temperature spin and charge transport properties of SLG device A. (a) Differential d$V$/d$I\times A$ curves of spin injection contact for as-fabricated device (black solid curve) and after each oxygen treatment. (b) Graphene resistance vs gate voltage. The inset is a close-up at negative gate voltages showing the onset of a second CNP at larger negative gate voltages as indicated by the increase of the resistance. (c) Spin resistance and (d) spin lifetime vs carrier density.}
	\label{fig:fig3}
\end{figure}

All contacts of the as-fabricated SLG device A are low resistive with $R_cA<\unit{0.1}{\kilo\ohm\micro\meter\squared}$. The respective (d$V$/d$I)\times A$ curve of the spin injection contact is shown in Fig.~\ref{fig:fig3}(a) (black curve). The gate dependent graphene resistance $R$ shows a single CNP at \unit{-2}{\volt} with only a small electron-hole asymmetry (black curve in Fig.~\ref{fig:fig3}(b)). As expected, spin transport through these as-fabricated transparent contacts exhibits small spin signals (Fig.~\ref{fig:fig3}(c)) and short spin lifetimes (Fig.~\ref{fig:fig3}(d)) with the latter being even below \unit{50}{\pico\second} at all carrier densities. Next, the device was kept in dry air for \unit{72}{\hour} with two subsequent exposures to pure oxygen at room temperature for \unit{39}{\hour} and \unit{102}{\hour}. The $R_cA$ values increase after each treatment (Fig.~\ref{fig:fig3}(a)) demonstrating the gradual oxidization of the graphene/MgO/Co interface. However, $dV/dI\times A$ curves of all contacts as a function of $I_{DC}$ remain completely flat with $R_cA<\unit{1}{\kilo\ohm\micro\meter\squared}$ indicating transparent contacts even after the last oxygen treatment [blue curve in Fig.~\ref{fig:fig3}(a)].
Nevertheless, we observe a clear increase of both the spin signal and the spin lifetime after each treatment showing that devices with larger $R_cA$ values exhibit enhanced spin properties. The changes in charge transport (Fig. 3(b)), on the other hand, are not as strong. After air exposure (red curve in Fig.~\ref{fig:fig3}(b)), the electron-hole asymmetry only slightly increases. After the first oxygen treatment the resistance shows a minimum at $\sim-40$~V which slightly moves to smaller negative gate voltages after the second oxygen treatment (see also inset in Fig.~\ref{fig:fig3}(b)). The minimum indicates the existence of a second CNP below $-80$~V, which is not accessible with our devices. Finally, we observe a weak \textit{p}-doping of the graphene after oxygen exposure (Fig. 3(b)) which we attribute to oxygen.\cite{doi:10.1021/nl1029607, NanoLett.SatoTakaiEnoki.2011}

The strong enhancement of the spin properties after each oxygen treatment together with the simultaneous increase of the $R_cA$ product is in agreement with our previous study\cite{PhysRevB.88.161405} and demonstrates that contact-induced spin dephasing is the bottleneck for spin transport in non-local spin-valves with low $R_cA$ values. While we discuss a second type of devices in the next section for which the spin properties can be enhanced even further, we first explore why the $R_cA$ values can be below 100~${\ohm\micro\meter\squared}$ as we typically measure for contacts with thin (2~nm) MgO barriers.

\begin{figure}[tb]
	\includegraphics{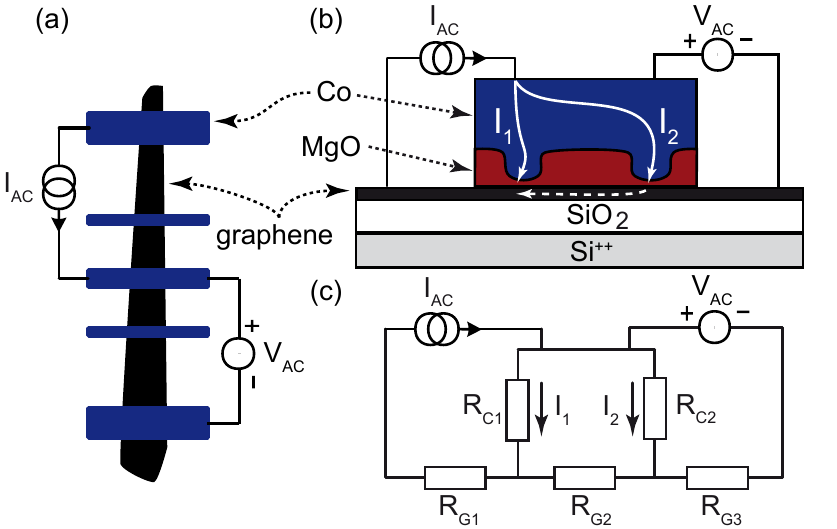}
	\caption{(Color online) (a) Schematic device layout with wiring for the measurement of the contact resistance. (b) Simplified schematic of the current flow through an inhomogeneous MgO barrier with conducting pinholes. (c) Equivalent circuit for (b).}
	\label{fig:fig4}
\end{figure}

As seen in the optical image of device B in Fig.~\ref{fig:fig5}(a) as well as in the schematic layout in Fig.~\ref{fig:fig4}(a) all electrodes can electrically be contacted from both sides of the graphene flake. This allows for a 4-terminal measurement of the contact resistance as illustrated in Figs.~\ref{fig:fig4}(a) and (b). As explained above, the island growth of MgO on graphene yields rather rough oxide barriers which favor pinholes for small nominal layer thicknesses. For simplicity, we show two conducting pinholes in the device cross-section in Fig.~\ref{fig:fig4}(b), where one is located near the current contact (left) while the other one is located near the voltage contact (right). The current which is driven from the left contact will primarily flow through both conducting pinholes with the total current $I_{\text{AC}}=I_1+I_2$. Because of the small resistance of  the metallic Co layer, we assume that the whole metallic electrode is one equipotential surface. However, this assumption does not hold for the underlying graphene as long as the contact resistance is of the same order as the graphene resistance. This case is very likely for our low resistive, transparent contacts and the pinhole connecting graphene part has to be treated individually when considering an equivalent circuit as in Fig.~\ref{fig:fig4}(c), where we assume a series of resistors for both the contact-covered and contact-free graphene parts of the device.

In this simplified equivalent network the real contact resistance $R_{\text{c,real}}$ is given by the two resistances $R_{\text{c1,2}}$ of the pinholes which are connected in parallel. Applying fundamental circuit laws gives:
\begin{equation}
\frac{R_{\text{c,meas}}}{R_{\text{c,real}}} = \frac{V_{\text{AC}}/I_{\text{AC}}}{R_{\text{c,real}}} = \frac{R_{\text{c1}}+R_{\text{c2}}}{R_{\text{c1}}+R_{\text{c2}}+R_{\text{G2}}}.
\end{equation}
We conclude that the measured contact resistance $R_{\text{c,meas}}$ is underestimated as long as $R_{\text{c,real}}$ is comparable to the graphene resistance $R_{\text{G2}}$ underneath the electrode. Of course, this model is simplified and gets much more complicated in two-dimensions with more pinholes. Nevertheless, it may explain qualitatively the unexpected small contact resistances in the as-fabricated devices with thin MgO barriers of up to \unit{2}{\nano\meter} thickness.

\subsection{BLG device}

\begin{figure}[tb]
	\includegraphics{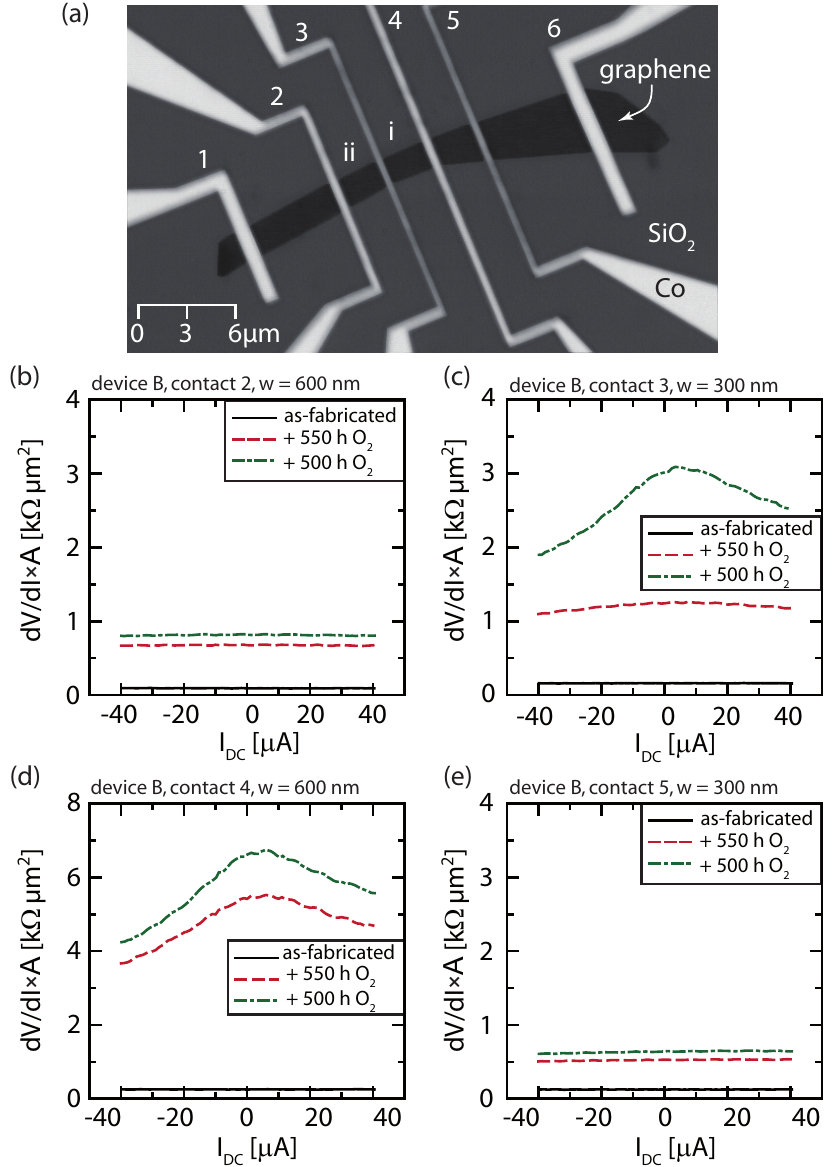}
	\caption{(Color online) (a) Contrast enhanced optical microscope image of spin transport device B (BLG). (b)-(e) Differential d$V$/d$I\times A$ curves of the inner contacts 2 to 5 for the as-fabricated state (black solid lines) and after each oxygen treatment. Only contacts 3 and 4 turn high resistive during oxygen treatments while contacts 2 and 5 remain transparent as seen by the flat d$V$/d$I\times A$ curves.}
	\label{fig:fig5}
\end{figure}

Fig.~\ref{fig:fig5}(a) shows an optical microscope image of the BLG device B. We enhanced the contrast significantly to better visualize the graphene flake. Similar to device A, the $R_cA$ values in the as-fabricated device B are well below $\unit{1}{\kilo\ohm\micro\meter\squared}$ [black solid curves in Figs.~\ref{fig:fig5}(b)-(e)].  After two oxygen treatments for $\unit{550}{\hour}$ and subsequent $\unit{500}{\hour}$ at room temperature two of the contacts (number 3 and 4 in Figs.~\ref{fig:fig5}(c) and (d), respectively) became high resistive and their differential d$V/$d$I$-curves are not flat anymore but rather show a cusp-like behaviour indicating the gradual transition to tunneling barriers. Interestingly, contacts 2 and 5 do not show such a pronounced development. Their d$V$/d$I\times A$ curves remain flat and the $R_cA$ values do not exceed $\unit{1}{\kilo\ohm\micro\meter\squared}$. It might be expected that contacts with smaller widths can be oxidized more rapidly from the sides but it becomes obvious from Fig.~\ref{fig:fig5} that the change in contact resistance does not depend on the electrode width as one of the broader electrodes ($w=600$~nm, Figs.~\ref{fig:fig5}(b) and (d)) and one of the narrow electrodes ($w=300$~nm, Figs.~\ref{fig:fig5}(c) and (e)) remains transparent while the respective other electrode turns towards tunneling contacts. Therefore, we assume that the coupling between graphene and MgO/Co electrodes must somehow differ between the electrodes of the same width.

\begin{figure}[tb]
\centering	
\includegraphics{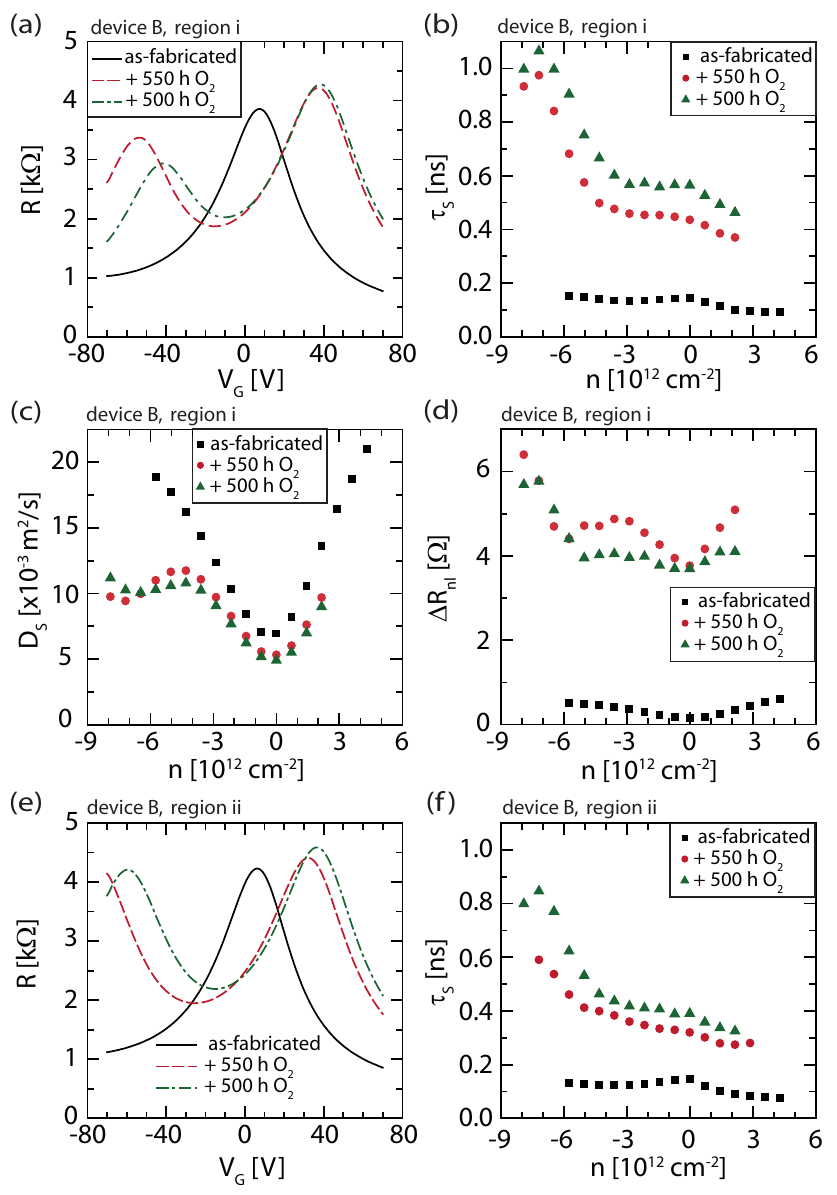}
	\caption{(Color online)  Room temperature spin and charge transport properties of BLG device B measured in region i for (a) to (d) and in region ii for (e) and (f) (regions are defined in Fig.\ref{fig:fig5}(a)). (a) Graphene resistance vs gate voltage. There is only one CNP in the as-fabricated device while a second contact-induced CNP appears at negative gate voltages after the first oxidization step. (b) Spin lifetime vs carrier density. Oxidization of contacts results in significant increase of $\tau_s$ near left CNP. (c) Spin diffusion coefficient vs carrier density. The minima near the CNPs are typical. (d) Spin signal vs carrier density. (e) Graphene resistance vs gate voltage in region ii with one high (contact 3) and one low (contact 2) resistive contact. In contrast to (a) the maxima of the left CNPs appear at larger negative  gate voltages. The increase of the respective spin lifetimes by oxygen treatments in (f) are less pronounced than in region i.}
	\label{fig:fig6}
\end{figure}

We first focus on charge and spin transport across the inner region of the device between electrodes 3 and 4 (region i) which both turn non-metallic after the oxygen treatments. For gate dependent charge transport measurements of the graphene resistance we send the current through contacts 1 and 6 and measure the voltage drop between electrodes 3 and 4. As seen in Fig.~\ref{fig:fig6}(a) the second contact-induced CNP has completely developed after the first oxidation step with its maximum at -53~V. Simultaneously, the respective contacts are now high resistive (red curves in Figs.~\ref{fig:fig5}(c) and (d)), which is both in agreement with our previous study where devices with as-fabricated high resistive contacts with $R_cA>\unit{1}{\kilo\ohm\micro\meter\squared}$ also exhibit a fully developed contact-induced second CNP.\cite{PhysRevB.88.161405} With further increase of $R_cA$ after an additional $\unit{500}{\hour}$ of oxygen treatment, we furthermore observe a decrease of the gate voltage separation between both CNPs. The right CNP which can be attributed to the contact-free graphene part in region i \cite{PhysRevB.88.161405} shifts from $V_{\text{G}}=\unit{7}{\volt}$ to $V_{\text{G}}=\unit{38}{\volt}$ most likely because of \textit{p}-doping by the oxygen.\cite{doi:10.1021/nl1029607, NanoLett.SatoTakaiEnoki.2011} Why this shift is far more pronounced compared to device A (Fig.~\ref{fig:fig3}(b)) is still an open question. We note, however, that the magnitude of \textit{p}-doping shows strong device-to-device variations and seems not to be related to the number of graphene layers (SLG or BLG). We guess that impurities and different couplings to the underlying substrate might play an important role for this behavior.\cite{doi:10.1021/nl1029607}

The influence of the oxygen treatments on the spin transport properties is shown in Figs.~\ref{fig:fig6}(b)-(d). As expected for $R_cA<\unit{1}{\kilo\ohm\micro\meter\squared}$ (as-fabricated device) both the overall spin lifetimes $\tau_s$ and the spin signals ${\Delta R_\text{nl}}$ are rather small (Figs.~\ref{fig:fig6}(b) and \ref{fig:fig6}(d)). However, the increase of $R_cA$ during oxygen treatment comes along with a significant increase of $\tau_s$ and ${\Delta R_\text{nl}}$ by a factor of more than 7 yielding ns spin lifetimes for gate voltages near the left contact-induced second CNP. Both, the spin diffusion coefficient $D_\text{s}$ and ${\Delta R_\text{nl}}$ (Figs.~\ref{fig:fig6}(c) and \ref{fig:fig6}(d)) show minima at each CNP while $\tau_s$ becomes maximal at those. A possible explanation for these findings is given in Sec.~\ref{oxidation_mechanism}. The results clearly show that the appearance of the contact-induced CNP strongly modifies the density dependent spin transport parameters.

Device B also allows studying spin transport through region ii for which only the right inner contact 3 turns high resistive after oxygen treatment (see Fig.~\ref{fig:fig5}(c)) while the other inner contact 2 remains metallic (see Fig.~\ref{fig:fig5}(b)). For the as-fabricated case, both the gate dependent graphene resistance (Fig.~\ref{fig:fig6}(e)) and the density dependent spin lifetime (Fig.~\ref{fig:fig6}(f)) are almost identical to region i. However, the overall changes during oxygen treatments are less developed in region ii which is seen by the smaller maximum spin lifetime in Fig.~\ref{fig:fig6}(f) (green triangles) from spin transport and the smaller gate voltage separation of the CNPs from charge transport (Figs.~\ref{fig:fig6}(e)). We obtain identical results when switching the current and spin diffusion directions between contacts 2 and 3 (not shown) demonstrating that spin and charge transport properties strongly depend on the electronic structure of both injection and detection contacts.

There is another interesting finding. We have never observed more that one contact-induced CNP in any of our devices nor have we seen any shoulders. This is also expected as long as all contacts of the same device have similar characteristics. It might be, however, surprising in devices with both low and high resistive contacts as in region ii of device B. A possible explanation might be inhomogeneous oxide barriers which may result in spatially inhomogeneous electrode-to-graphene interaction. If the variation in interaction under each electrode is larger than their averaged difference, then the large full width at half maximum values of the two contact-induced CNPs may lead to one broad resistance peak. But there might be also other contributions such as local electric fields between the contact-covered and contact-free graphene parts which have to be considered here.\cite{PhysRevB.79.245430}

Finally, we want to point out that the maximum resistances from the contact-induced CNPs are surprisingly large. As seen in Fig.~\ref{fig:fig5}(a) there is a significant difference in the areas of contact covered and uncovered graphene. From the \unit{3}{\micro\meter} spacings between the inner electrodes 2 to 5 (measured from center-to-center between respective neighboring electrodes) only \unit{0.45}{\micro\meter} are covered by contacts. But the measured maximum graphene resistance under the contacts (contact-induced left CNP) is comparable to the graphene resistance between the contacts (right CNP) (see Figs.~\ref{fig:fig6}(a) and \ref{fig:fig6}(e)). This implies that the graphene resistivity under the contacts is significantly larger than for graphene parts in between the contacts. This larger resistivity might be caused by distortions of the graphene lattice under the electrodes. Especially at pinholes or very thin parts of the oxide barrier the contact interaction of graphene with Co may yield $sp^3$ type hybridization with the underlying graphene \cite{PhysRevLett.101.026803,PhysRevB.87.075414}, which may result in larger resistivities. This notion will also be important in Sec. \ref{oxidation_mechanism}.

\subsection{Comparison to previous data}

\begin{figure*}[t]
	\includegraphics{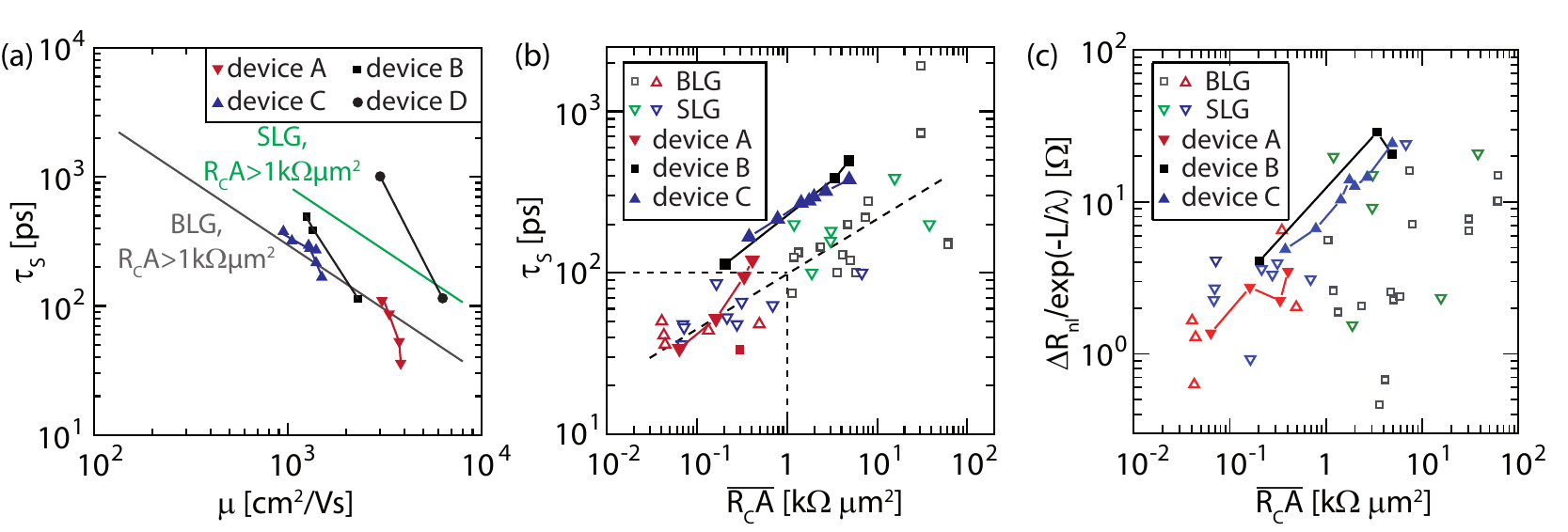}
	\caption{(Color online) (a) Spin lifetime vs electron mobility at $n=\unit{1.5 \times \power{10}{12}}{\centi\meter\rpsquared}$ at RT for devices A to D. The lines are taken from Ref.~\onlinecite{PhysRevB.88.161405} and illustrate the measured $1/\mu$ dependence in as-fabricated SLG and BLG devices with large $R_cA$ values ($R_cA>\unit{1}{\kilo\ohm\micro\meter\squared}$). (b) Spin lifetime vs averaged contact resistance area product $\overline{R_cA}$ of respective injection and detection electrode at an electron density of $n=\unit{1.5 \times \power{10}{12}}{\centi\meter\rpsquared}$ at RT. For comparison we included all previous measurements on as-fabricated SLG and BLG devices from Ref.~\onlinecite{PhysRevB.88.161405}. (c) Normalized non-local spin resistance vs. averaged resistance area product. While there is a clear increase of the spin resistance after contact oxidization, there is no trend in previous results on as-fabricated devices which are included from Ref.~\onlinecite{PhysRevB.88.161405}.}
	\label{fig:fig7}
\end{figure*}

It is interesting to evaluate the dependence of $\tau_{\text{s}}$ on the electron mobility $\mu$ as shown in Fig.~\ref{fig:fig7}(a). Here, the mobility from the right CNP was calculated at an electron density of $n=\unit{1.5 \times \power{10}{12}}{\centi\meter\rpsquared}$ by a linear fit $\mu =(1/e)(\Delta\sigma/\Delta n)$ to the conductivity. In addition to devices A and B we have also included results from secondary SLG (device D) and BLG devices (device C). In all devices there is a striking decrease of $\mu$ after each oxygen treatment while $\tau_{\text{s}}$ increases at the same time. The strongest increase of $\tau_{\text{s}}$ is obtained for device D (circles) with $\tau_{\text{s}}$ exceeding 1~ns. While all other devices were treated as described above, device D was stored partially in dry air and in vacuum and was remeasured 2 years after fabrication. Despite these different storage and treatment procedures there is surprisingly a similar slope in the $\tau_{\text{s}}$ vs $\mu$ dependence.  As discussed in Ref.~\onlinecite{PhysRevB.88.161405}, the superposition of the emerging contact-induced left CNP with the right CNP from the graphene part between the electrodes changes the slope $\Delta\sigma/\Delta V_G$ at $n=\unit{1.5 \times \power{10}{12}}{\centi\meter\rpsquared}$ ($V_G-V_D\approx20$~V). Next to the oxygen doping of graphene \cite{NanoLett.SatoTakaiEnoki.2011} we therefore attribute the overall decrease in $\mu$ to the smaller voltage separation between both CNPs after each oxygen/air treatment which results in a smaller slope $\Delta\sigma/\Delta V_G$ and thus explains the drop in the extracted $\mu$.

Another important finding is that the $\tau_{\text{s}}$ vs $\mu$ dependence under oxygen treatment does not follow the $1/\mu$ dependencies which we observed in as-fabricated SLG and BLG devices with $R_cA>\unit{1}{\kilo\ohm\micro\meter\squared}$ which we included in Fig.~\ref{fig:fig7}(a) from our previous studies as a green and a grey line, respectively. For easier comparison, we did not include the actual data point of the old devices, which can, however, be found in Ref.~\onlinecite{PhysRevB.88.161405}. Although the second CNP is also distinctly visible in as-fabricated samples with large $R_cA$ values exhibiting long spin lifetimes (see also Fig.~\ref{fig:fig1}(a)) we can exclude from our present study that the contact-induced CNP alone can account for the observed $1/\mu$ dependence of $\tau_{\text{s}}$ in Ref.[\onlinecite{PhysRevB.88.161405}].

One of the key results in our previous work was the increase of the spin lifetimes with the $R_cA$ values highlighting the role of contacts to the overall spin dephasing in as-fabricated devices.\cite{PhysRevB.88.161405} But as discussed for device A and B the dominating influence of the contacts on the spin lifetime also becomes apparent in the present study. We therefore calculated the average value of the contact resistance area products $\overline{R_cA}$ of injection and detection electrodes and combine results from devices A to C with the data from Ref.~\onlinecite{PhysRevB.88.161405} in Fig.~\ref{fig:fig7}(b). Device D got damaged before the $R_cA$ values could be measured. The observed increase of $\tau_s$ during the oxygen treatment is in good agreement with our previous results indicating that similar spin dephasing and spin relaxation processes limit spin transport for devices with lower $\overline{R_cA}$ values. We note that there is still a large scatter in the respective values for as-fabricated devices demonstrating that next to the contact characteristics, other sources for spin dephasing and spin relaxation have to be considered which can yield device-to-device variations such as electronic defects in the underlying wafer or charged impurities or defects in the graphene flake. We expect, however, that these contributions do not significantly change during oxygen treatment. This notion is supported as the slope of the $\tau_s$ vs $\overline{R_cA}$ dependence of the oxygen treated devices completely follows the general trend of all other devices.

As it was discussed for devices A and B in Figs.~\ref{fig:fig3}(c) and \ref{fig:fig6}(d) the non-local spin resistance $\Delta R_{\text{nl}}=\Delta V_{\text{nl}}/I$ also strongly increases with increasing $R_cA$ values. This might also be expected when considering the predicted role of oxide barriers on the conductivity mismatch problem for spin transport.\cite{PhysRevB.64.184420,PhysRevB.62.R16267}

In non-local 4-terminal spin-valve measurements $\Delta R_{\text{nl}}$ is given by:
\begin{equation}
\Delta R_{\text{nl}}=R_0 \frac{\rho\lambda}{w}\text{exp}\left( -L/\lambda \right),
\label{equationDRnl}
\end{equation}
where $\rho$ is the sheet resistance of the graphene flake, $\lambda=\sqrt{D_S \tau_S}$ the spin diffusion length, $w$ the width of the graphene flake, $L$ the distance between injector and detector electrode and $R_0$ a factor that represents the contact characteristics. Different expressions for $R_0$ are derived theoretically and have been used for analysis of spin transport experiments.\cite{ISI000249789600001,Nature.416.713-716,Takahashi2003,Tombros2007,PhysRevLett.109.186604}

To account for device-to-device variations in injector/detector distance $L$ we plot the spin resistance $\frac{\Delta R_{\text{nl}}}{\text{exp}\left( -L/\lambda \right)}$ values of devices A to C in Fig.~\ref{fig:fig7}(c), where we additionally included results from all previous as-fabricated devices presented in Ref.[\onlinecite{PhysRevB.88.161405}] for comparison. While we could not observe any systematic dependence in previous data (see open symbols) we see a clear increase of the spin resistance with increasing $\overline{R_cA}$ in all present devices (see full symbols in Fig.~7(c)). Although these seemingly inconsistent results  might be surprising, they again demonstrate that spin and charge transport parameters are governed by many microscopic parameters and details which are currently not completely unveiled.

\section{oxygen doping of the graphene flake}
\label{oxygen_doping}

\begin{figure}[b]
	\includegraphics{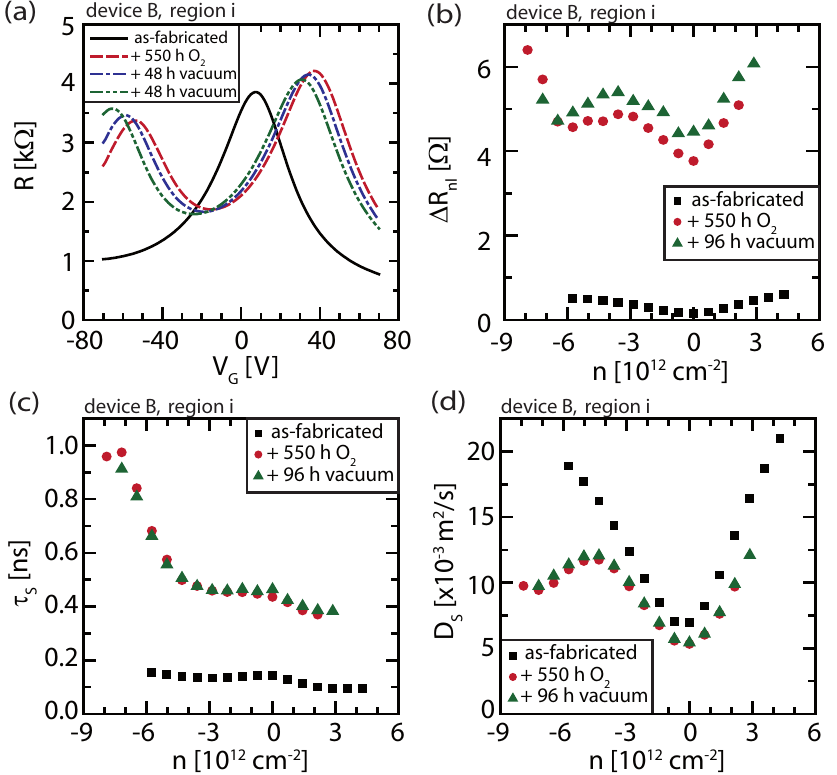}
	\caption{(Color online) Room temperature spin and charge transport properties of BLG device B measured in region i. (a) Graphene resistance vs gate voltage. After the first oxidization step (red curve) the device was twice kept in vacuum for 48~h which results in a subsequent shift of the left CNP towards larger negative gate voltages. (b) Spin signal vs carrier density. Vacuum storage does neither change the density dependent spin lifetimes (c) nor the spin diffusion coefficients (d).}
	\label{fig:fig8}
\end{figure}

In this section we first address the question if the measured spin lifetime is also influenced by the oxygen doping of the graphene flake. In this context, it is important to note that doping of graphene by oxygen occurs on much shorter time scales than our treatment time of \unit{500}{\hour} and saturates eventually.\cite{doi:10.1021/nl1029607} Because of this saturation, the position of the right CNP which results from the contact-free graphene part between the contacts is almost the same for both oxidation steps in device B (Figs.~\ref{fig:fig6}(a) and \ref{fig:fig6}(e)). Therefore, the increase in spin lifetime in both regions after the second oxygen treatment has to be related to the change in contact resistance and not to oxygen doping. We also repeated the measurements of device B after the first oxidation step twice after keeping the device in vacuum for 48~h in each case. Apparently, part of the less bound oxygen gets released as both CNPs are shifting in gate voltage (Fig.~\ref{fig:fig8}(a)). In contrast, the differential contact resistance (not shown), the spin lifetime (green triangles in Fig.~\ref{fig:fig8}(c)) and the spin diffusion coefficient (green triangles in Fig.~\ref{fig:fig8}(d)) do not change. Accordingly, oxygen doping of the graphene flake does not considerably influence the spin lifetime at room temperature. Only the spin signal is slightly increased after partial oxygen loss in vacuum (Fig.~\ref{fig:fig8}(b)).

\begin{figure}[tb]
	\includegraphics{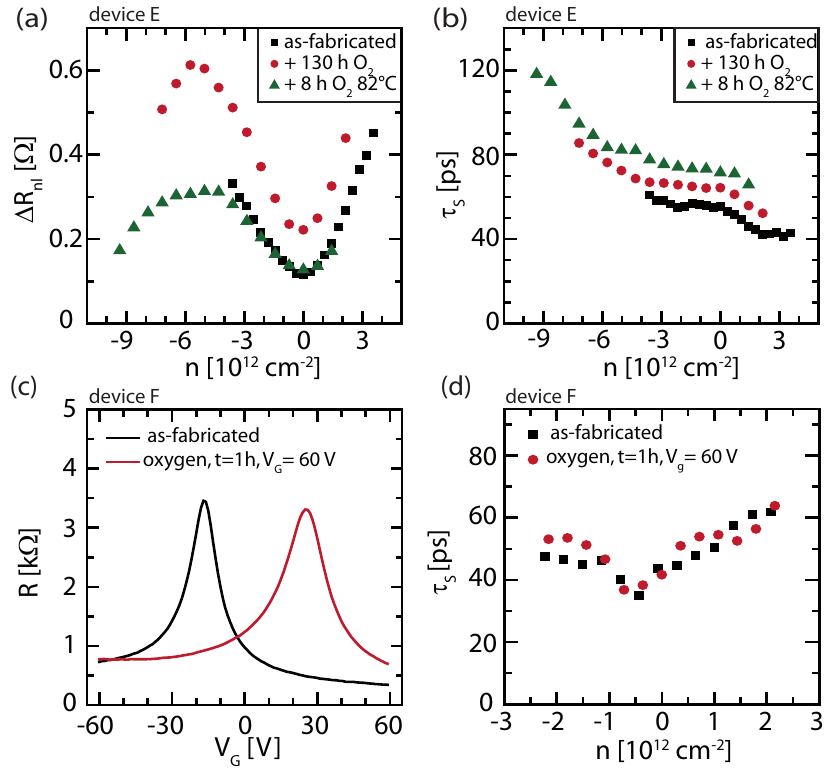}
	\caption{(Color online)  (a) Charge carrier density dependent non-local spin-valve signal and (b) spin lifetime of device E (SLG) after oxygen treatments at higher temperatures. The increase in lifetime combined with the decrease of the spin signal after the second treatment illustrates that different mechanisms influence these spin transport parameters. (c) Gate dependent resistance of device F before and after a short oxygen exposure using a gate electric field. (d) The oxygen doping causes no obvious effect on the spin lifetime.}
	\label{fig:fig9}
\end{figure}

It is important to note that the spin signal and the spin lifetime can develop differently, in particular if the oxygen treatment is carried out at higher temperatures. The heating of the device during oxidation was an attempt to accelerate the oxidization process. A disadvantage of this approach is that the device can easily be damaged. We found that annealing temperatures larger than \unit{100}{\degreecelsius} often result in strongly reduced or completely vanishing spin signals although the electrical contact over the graphene flake is rather robust during that treatment. In Figs.~\ref{fig:fig9}(a) and \ref{fig:fig9}(b) we plot the change in spin resistance and spin lifetime in a separate SLG device E after two oxygen treatments, respectively. In agreement with all the other devices, room temperature oxidization for 130~h yields an increase in both $\Delta R_{nl}$ and $\tau_s$. In contrast, $\Delta R_{nl}$ decreases to the initial values after the second oxygen treatment at \unit{82}{\degreecelsius}. Notably, the spin lifetime increases after each treatment demonstrating that different mechanisms influence both spin transport parameters.

To back-up our claim that the increase in spin lifetime is not caused by oxygen doping of graphene, we finally show results for device F (SLG) in Figs.~\ref{fig:fig9}(c) and \ref{fig:fig9}(d). This device has transparent contacts and was exposed to oxygen only for one hour but with an applied gate voltage of $V_{\text{G}}=\unit{60}{\volt}$. An applied gate electric field during oxygen exposure is known to enhance the doping efficiency of oxygen.\cite{NanoLett.SatoTakaiEnoki.2011} Accordingly, the CNP of the device is significantly shifted by \unit{42}{\volt} to the right. On the other hand, the exposure time is too short to effectively change the MgO/graphene interface as seen by the single CNP without even an onset of the contact-induced CNP at negative gate voltages. Despite the significant carrier doping by oxygen we therefore neither observe any notable change in the spin signal (not shown) nor in the spin lifetime [Fig.~\ref{fig:fig9}(d)].

\section{possible oxidation mechanism of the graphene/MgO/Co-interface}
\label{oxidation_mechanism}

In this section we discuss possible oxidization mechanisms at the graphene/MgO/Co interface during oxygen treatment and develop implications on spin and charge transport properties. Previously, it was reported that the interface between metallic electrodes made out of Al, Co and Cu to exfoliated graphene can be belatedly oxidized by exposing the as-fabricated devices to air for a sufficient period of time.\cite{Nam2012,nouchi:253503,Malec2011,Miyazaki2012,APEX.1.034007} It has been argued that the weak bonding between graphene and these metals may lead to sideways diffusion of oxygen along the metal-to-graphene interface.\cite{Nam2012,Malec2011} This notion is underpinned by time-resolved low energy electron microscopy studies in which the oxidization of a Ru(0001)-single crystal covered by graphene grown by chemical vapor deposition is shown.\cite{Sutter2010}

If oxygen diffuses from the edges of the contacts it is obvious to assume that oxidization of wider contacts is less efficient and thus takes longer time than for narrower contacts. But as discussed in Sec.~\ref{oxygen_treatments}B we observe significant variations in the oxidization efficiency of even equally dimensioned electrodes. These might also result from different couplings of the graphene flake to the underlying substrate.\cite{doi:10.1021/nl1029607} It was reported that exfoliated graphene flakes do not necessarily completely follow the surface roughness of a $\textrm{SiO}_2$ substrate\cite{Ishigami2007,Cullen2010} and therefore different diffusion behaviors of water and gases along the graphene-to-$\textrm{SiO}_2$ interface have been proposed.\cite{doi:10.1021/nl1029607,raey} The question arises whether these spatial variations in the coupling to the substrate may also influence the post-oxidization of the MgO barriers and can explain the observed differences in the efficiency of the oxygen treatments presented in this work.

It is likely that spatially varying lithography residues between graphene and the deposited electrodes may significantly influence oxygen diffusion along the interface. Furthermore, a high resolution TEM  image (not shown) of one of our as-fabricated devices shows the polycrystalline structure of the MgO barrier which is not surprising when considering the growth mode described in Sec.~\ref{fabrication}. Such a polycrystalline barrier has defects and grain boundaries, which enhance the reactivity of MgO$_x$ and may also support diffusion of oxygen.\cite{Liu1998} Moreover, only a rather small amount of additional oxygen is needed to reach the graphene/MgO/Co interface in order to compensate possible oxygen vacancies in the MgO$_x$ barrier and to oxidize conducting pinholes which both improve the barrier quality.

\begin{figure*}[t]
	\includegraphics{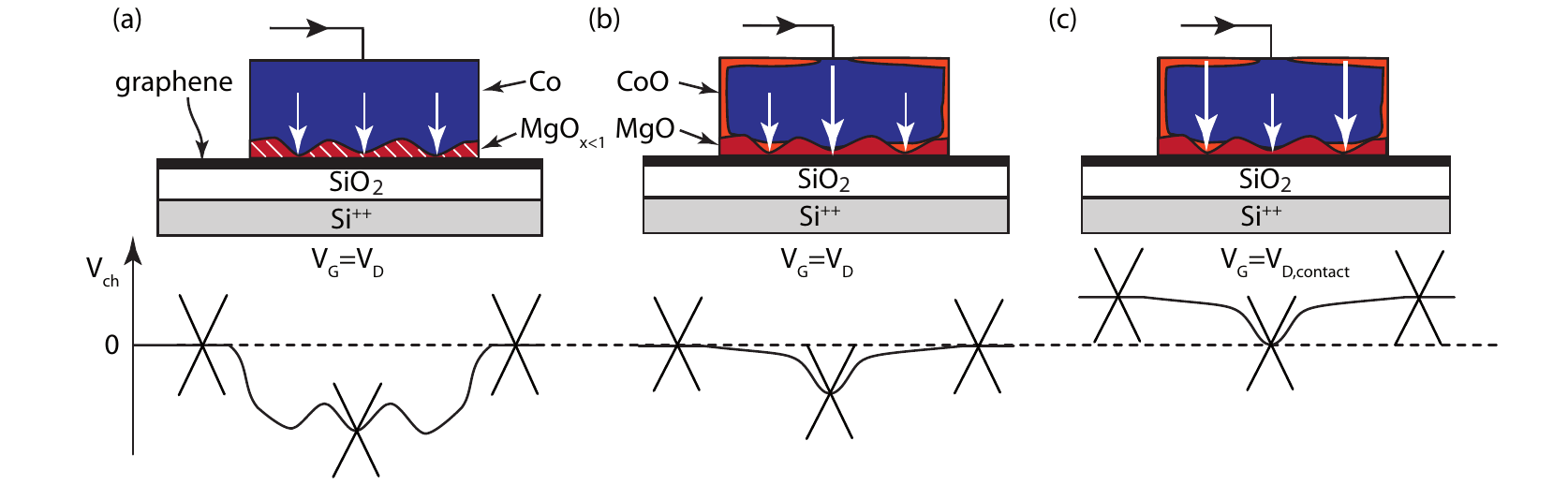}
	\caption{(Color online) Illustration of modifications in the MgO oxide barrier, changes in the current distributions (white arrows) and doping profiles of the electrostatic potential during the oxygen treatments. The situation is shown for the as-fabricated condition in (a) and after oxygen treatment at different gate voltages in (b) and (c). See text for further explanations.}
	\label{fig:fig10}
\end{figure*}

The existence of pinholes in the as-fabricated devices not only explains the surprisingly small measured contact resistances (see Sec.~\ref{oxygen_treatments}A) but also the absence of the contact-induced second CNP and even the very small spin lifetimes. At the position of pinholes the graphene is in close contact with Co atoms. It has been theoretically predicted that Co gets chemisorbed on graphene and modifies its band structure to a large extent.\cite{PhysRevLett.101.026803} Angle-resolved photoelectron spectroscopy of CVD-grown graphene on Co substrates demonstrates that the Dirac cone is significantly shifted into the valence band and that the $\pi^*$ band of graphene hybridizes with the 3\textit{d} bands of Co near the Fermi level.\cite{Varykhalov2009,Varykhalov2012} On the one hand, these pronounced hybridization states can explain why the Fermi level gets pinned and prohibits gate voltage tunability in graphene underneath transparent Co electrodes. But on the other hand it is important to emphasize that charge and spin transport takes place through these hybridized states. Therefore spin scattering induced by the hybridized 3\textit{d} states may explain the very short spin lifetimes. In this context it would be interesting to explore whether the pinholes may behave like spin hot spots which were recently proposed for hydrogen adatoms on graphene as a very efficient source for resonant spin scattering yielding short spin lifetimes.\cite{PhysRevLett.112.116602}

By our oxygen treatments the pinholes are electrically pinched off and the overall thickness of the oxide barrier increases which leads to a weakening of the Co/graphene interaction. As a result both the \textit{n}-doping and the density of hybridized 3\textit{d} states near the Fermi level get gradually diminished. This eventually leads to the appearance of the second charge neutrality point at large negative gate voltages and the increase in spin lifetime. Accordingly both Dirac peaks should be merging as soon as a continuous oxide barrier of sufficient thickness leads to a decoupling between graphene and Co assuming that the MgO does not cause any additional doping.

Based on these arguments we now propose a model which explains the strong increase of the spin lifetime at the contact-induced charge neutrality point of device B [Figs.~\ref{fig:fig6}(a) and \ref{fig:fig6}(b)]. We assume that in the as-fabricated condition the MgO barrier is oxygen deficient (MgO$_x$ with $x<1$ according to Sec.~\ref{fabrication}) and contains conducting pinholes over the whole contact area as illustrated in Fig.~\ref{fig:fig10}(a). Accordingly, the applied current will primarily flow through the areas of the pinholes (see white arrows). The doping profile of the electrostatic potential in the graphene is illustrated for the case that the gate voltage is tuned to the CNP of the bare graphene parts on the left and on the right side of the electrode (see for example Fig.~\ref{fig:fig10}(a) for as-fabricated device (black curve) with $V_{\text{G}}=V_{\text{D}}=\unit{7}{\volt}$). The graphene under the electrode and in particular at the positions of the pinholes is highly \textit{n}-doped because of the interaction with the Co. While the Fermi level of the outer graphene parts can be tuned by the gate voltage, it remains pinned under the contact due to the hybridized states [see also schematic in Fig.~\ref{fig:fig1}(b)]. Therefore no pronounced gate dependence of the spin lifetime can be observed in Fig.~\ref{fig:fig6}(b). Following the discussion in the last section, the overall spin lifetime is also very short.

During oxygen treatments both the healing of oxygen vacancies in the MgO$_x$ layer and an additional oxidization of the Co may occur. The effective thickness of the oxide barrier most likely increases faster at the sides than in the center of the electrodes, leading to the inhomogeneous oxide barrier thickness as illustrated in Figs.~\ref{fig:fig10}(b) and \ref{fig:fig10}(c). In Fig.~\ref{fig:fig10}(b) we first discuss the situation with the gate voltage tuned to the CNP of the bare graphene between the electrode, i.e. $V_{\text{G}}=V_{\text{CNP}}=\unit{37}{\volt}$ as seen in Fig.~\ref{fig:fig6}(a) by the red dashed and green dash-dotted line. The thicker the oxide barrier becomes the more the doping of the graphene under the electrode approaches the one of the outer bare graphene parts. Only in the middle region of the electrode, where the oxidization is less advanced, there remain strong \textit{n}-doping from the Co. Due to the pinhole the barrier resistance is smallest there and most of the applied current will now flow through this part (see large white arrow in Figs.~\ref{fig:fig10}(b)). But as the oxide barrier at this point is slightly thicker than in the as-fabricated device and also because of small contributions of the current over the thicker parts of the oxide barrier, the spin lifetime becomes longer than in the initial device.

The crucial difference to the as-fabricated condition is that now the oxide barrier is thick enough to eliminate Fermi level pinning and to significantly reduce the \textit{n}-doping by the Co. Accordingly, it is now possible to tune the graphene carrier density to its CNP near the center pinhole. The corresponding doping profile is illustrated in Fig.~\ref{fig:fig10}(c) (see also Fig.~\ref{fig:fig6}(a) after the first oxygen treatment at $V_{\text{G}}=V_{\text{CNP,contact}}=\unit{-53}{\volt}$).

Now a key result from Sec.~\ref{oxygen_treatments}B gets important: the maximum resistance at both CNPs is comparable although there is a significant difference in the area of the contact-covered and contact-free graphene parts as seen in Fig.~\ref{fig:fig5}(a). This implies that the resistivity of the contact covered graphene must be very large near the contact-induced CNP. Therefore, the combination of oxide barrier resistance and graphene resistance can now also favor currents over the thicker parts (see large arrows in Fig.~\ref{fig:fig10}(c)) of the oxide than over the thinner part in the middle of the electrode (see small arrow in Fig.~\ref{fig:fig10}(c)). This notion is further supported as the graphene resistance under the thicker oxide is now smaller due to the larger gate-induced charge carrier density as illustrated in the doping profile in Fig.~\ref{fig:fig10}(c). On the other hand, it is very unlikely that spins that are injected over these thicker parts of the oxide diffuse into a graphene region of the thinner oxide barrier mainly because of its small density of states at the CNP. But it will be exactly these leftover inner pinhole regions in which spin dephasing and scattering remains strongest. The reduced interaction of the injected spins with these spin hot spots may explain the strongly suppressed contact-induced spin scattering at this gate voltage which allows for nanosecond spin lifetimes as shown in Fig.~\ref{fig:fig6}(b).

\section{conclusions}
\label{conclusions}

In summary, we successfully manipulated the graphene-to-MgO/Co interface in single and bilayer graphene spin-valve devices by successive oxygen treatments and achieved a gradual transition from small towards high contact resistance area products. With this the spin transport properties are enhanced significantly. We observe an increase in spin lifetimes and spin resistances by a factor of seven and achieve values of up to \unit{1}{\nano\second} at room temperature. Therefore the limiting factor of contact-induced spin dephasing is demonstrated without error-prone comparison between different devices.

Subsequent oxidization of the MgO spin injection barrier furthermore leads to the appearance of a contact-induced charge neutrality point in charge transport. We observe that this Dirac peak is strongly linked to gate dependent spin transport as seen by the increase of the spin lifetime. We gave a possible explanation for this behavior considering both the Co-to-graphene interaction and the spatially inhomogeneous spin and charge current paths through the MgO barrier. This explanation highlights one important aspect: Both, the spin and charge transport must simultaneously be understood on a microscopic level in order to unveil relevant spin scattering mechanisms from macroscopic Hanle measurements of the whole device. This was already partially addressed in a previous study that theoretically investigated the implications of different spin transport parameters of contact-covered and contact-free graphene regions on the extracted value of the spin lifetime by Hanle measurements.\cite{doi:10.1021/nl301050a} But the real situation is by far more complicated. For example, there will be $p-n$ junctions formed between the differently doped contact-covered and contact-free graphene regions with large local electric fields which might also play an important role for spin transport. We especially demonstrated that not only differences between the contact covered and bare graphene regions are important to understand the overall charge and spin transport properties of a device but also subtle variations within the contact regions which result from inhomogeneous oxide barriers and pinholes.

The research leading to these results has received funding from the DFG through FOR-912 and the European Union Seventh Framework Programme under grant agreement n${^\circ}$604391 Graphene Flagship.

%

\selectlanguage{english}

\end{document}